# New Bounds on Ω Baryons from Observational Big Bang Nucleosynthesis

Dimitar Sasselov *

*Harvard-Smithsonian Center for Astrophysics,*
*60 Garden Street, Cambridge, MA 02138, USA; e-mail: sasselov@cfa.harvard.edu*

Dalia Goldwirth

*Dept. of Physics and Astronomy, Sackler Fac. of Exact Sci.,*
*Tel-Aviv Univ., Tel-Aviv 69978, Israel; e-mail: dalia@wise7.tau.ac.il*

(July 6, 1994)

## Abstract

We re-examine the systematic errors in determination of the primordial helium abundance, $Y_P$. We find that the systematics are significantly larger than the statistical errors. The uncertainty in (the determination of) $Y_P$, is thus, larger than is currently claimed. Furthermore, most of the systematics lead to underestimate of $Y_P$. The new upper bound allows cosmological models with no non-baryonic dark matter in which $\Omega_{baryons} = \Omega_{BBN} = \Omega_{dyn}$.

## I. INTRODUCTION

The current limit on $\Omega_{BBN}$, the baryonic density parameter that matches big bang nucleosynthesis, is $\Omega_{BBN} < 0.06 \ h_{50}^{-2}$ [1](Where $h_{50}$ is the Hubble constant in units of 50 km $s^{-1}$ $Mpc^{-1}$.) This limit is in conflict with other observations of the density parameter $\Omega$; From inferred motions of virialized systems on small scales we get a dynamical estimate to $\Omega$ which is $\Omega_{dyn} = 0.1 - 0.2$ [2]. In order to satisfy the above BBN limit one has to invoke unknown non-baryonic dark matter. Theoretical prejudice, based on the inflationary scenario, favor a flat universe with $\Omega_{total} = 1$. However, the observed ratio of matter+gas to total mass in clusters [3] $\Omega_{baryons}/\Omega_{total} \geq 0.01 + 0.14 h_{50}^{-1.5}$ is in conflict with the above constraint on $\Omega_{BBN}$ for $\Omega_{total} = 1$. The primeval baryon isocurvature model (PBI) [4], with $\Omega_{total}$ being only baryonic between $0.1 - 0.2$, i.e. $\Omega_{total} = \Omega_{BBN}$, seems to fit observations better than any other cosmological models [5], except for one – the observation limit on $\Omega_{BBN}$. This uncomfortable low limit has provoked many attempts [6] to increase the bound on $\Omega_{BBN}$. Most of these trials have been unsuccessful [7], while, the few successful scenarios [8] [9] are based on drastic modifications of the standard picture.

---

*Hubble Fellow

In this paper we reexamine the observational basis for the constraints imposed on $\Omega_{BBN}$. Of all the BBN isotopes, $^7$Li is the easiest abundance to determine, however, it is very fragile which makes it hard to relate it to its primordial content. Just recently the $^7$Li abundance was increased by a factor of 7 [10] as a result of studies of both young and old population stars which require the inclusion of diffusion as well as rotational mixing in the stellar models [11]. This increase can be constrained by further detections of $^6$Li in old stars, but progress there is still slow. Independently, improvements in reaction rates and the neutron half-life [12] [1], in addition to recent correction in the $T_{\text{eff}}$-scale of halo stars [13] brought the $^7$Li abundance even further up. The new values for $^7$Li require higher values for the current $^4$He abundance for consistency.

Deuterium, can only undergo net destruction following the epoch of BBN. Because of its steep dependence on the ratio of baryons to photons, $\eta$, the D abundance could play a major role in studying BBN, if constrained more strongly. Unfortunately, current values of $D_P$ from interstellar medium determinations have a large scatter around D/H $\approx 1.5 \times 10^{-5}$ [14], while the Solar System measurements are limited by the need to know the solar $^4$He abundance [15] which has not been derived accurately yet. When the uncertainties are accounted for, the current lower limits on D and D+$^3$He allow our upper bound on the helium primordial abundance of $Y_P \leq 0.255$; however, a $Y_P$ higher than that runs into conflict with the presently estimated values of $D_P$ [16]. Quite recently two groups reported the possible detection of extragalactic D in the spectrum of a high-redshift QSO [17] [18]. They derive a deuterium abundance of D/H $\approx 25 \times 10^{-5}$ corresponding to $Y_P$ near 0.235, but caution that the observed feature has a high probability of being due to hydrogen absorption along the same line of sight. However, if the absorption is due to deuterium the derived abundance is so high, that it is in conflict with the solar system observations of D and $^3$He [19].

As a result of the recent uncertainties in the determination of $^7$Li and D, the abundance of $^4$He continues to play a key role in defining the observational constraints on $\Omega_{BBN}$. The uncertainties in the BBN observational limits are dominated by systematic errors which generally escape the attention of those who use $Y_P$ limits for theoretical constraints. During the last few years there has been substantial improvement in the atomic data for helium, as well as of the methods for solving detailed radiative transfer. In this paper we take advantage of both in order to evaluate the systematic errors in the derivation of the primordial helium abundances. Our goal here is to obtain a quantitative estimate of the systematic errors and new upper bounds on $Y_P$ and on $\Omega_{BBN}$, respectively. We expand on the implication of this new bound on various cosmological scenarios.

## II. THE $^4$HE LIMIT

The observational determination of $Y_P$ relies exclusively on the use of the optical recombination spectra of metal-poor, extragalactic ionized nebula – H II regions and H II galaxies [20] [21] [22]. The major advantage of using such H II regions is the theoretical simplicity of recombination radiation. In the absence of complicating effects, the He/H abundance ratio could be deduced from relative intensities of hydrogen and helium lines using theoretical effective recombination coefficients. The procedure of deriving the abundances from the observations consists of adopting a simple Case B radiative recombination theory [23] followed by a set of "corrections" which are applied to counter its inadequacy [21]. In Case B one assumes that the line photons (mostly resonance lines of abundant ions) are scattered so many times that their downward radiative transitions can effectively be omitted from consideration. The other extreme assupmtion, Case A, which is valid for most of the emission lines observed in nebulae, is that all emitted line photons escape and there is no radiative transfer problem. The real situation, however, is intermediate (Osterbrock [24], Ch.4), non-linear and dependents on the ions involved. The important sources of errors inherent to the H II galaxy method have been already identified and discussed (see *e.g.*, references [25] [26] and [21]). Most of these are of systematic nature and fall generally into the category of radiative transfer problems. Below is a list of some of these radiative transfer problems:

**(1)** The invalidity of Case B for $H$.

**(2)** The invalidity of Case B for $He\ I$.

**(3)** The amount of neutral helium.

All the listed effects, plus some we do not list (e.g., consideration of underlying stellar absorption), always lead to a systematic underestimate of $Y_P$ (see Skillman & Kennicutt [25]).

We examine the above systematic effects with the help of detailed radiative transfer calculations for H, He I & II. We solve simultaneously the coupled statistical equilibrium and transfer equations in plane-parallel illuminated slab geometry using the approximate $\Lambda$-operator method of Scharmer [27]. This is a full multi-level non-LTE calculation for the atomic models of H, He I, and He II. We take special care of the helium resonance lines transfer, e.g., He II L$\alpha$ 304 Å, which can be an important ionization source. In constructing the slabs we used photoionization equilibrium calculations with a mean escape probability formulation after Proga et al. [28] and similar to model #8 of Clegg & Harrington [29]. The helium atomic model is similar to that of Almog & Netzer [30], but contains some new atomic data from the Opacity Project database, the new term values from Drake [31], and the collisional rates evaluated by P.J. Storey from the cross sections of Berrington & Kingston [32](see Sasselov & Lester [33] for some details on the atomic data). We compare the results of our computation to the calculations using Case B analysis with the standard "corrections" for the same physical problem. The comparison gives us a *realistic* estimate of the systematics in the determination of $Y_P$ from the observed data.

We apply the results of our calculations of systematic effects to four recent values of $Y$ and $Y_P$ as derived by different observers: $Y_P$ by Pagel et al. [21] derived from He *vs*. O and He *vs*. N; $Y_P$ by Olive & Steigman [34] from a new set of low-metallicity H II regions observed and analyzed by Skillman et al. [22] and combined with the set of Pagel et al.(for both O and N); and the abundance determination in the most low-metallicity system – I Zw 18 by Skillman & Kennicutt [25].

We proceed by describing the specific problems and their overall contribution to the individual systematic errors for each case.

*[(1)] The invalidity of Case B for H:* Several processes complicate the radiative transfer of hydrogen in H II regions [35], [36], [37]. First, the low-metallicity H II regions have high enough electron temperatures that some of the observed H$\alpha$ flux may be due to collisional excitation. The rate depends on the ground-state population, $N_{H^o}$, which cannot be derived

from recombination theory alone. Worse still, the amount of neutral hydrogen is *unobservable* in H II regions, due to the lack of any spectral diagnostics. The ground-state population, $N_{H^0}$, can be obtained from detailed radiative transfer calculations with geometry constrained by high-resolution observations.

To demonstate the magnitude of this effect we consider NW H II region in I Zw 18 for which high resolution data is available. In this case we can calculate the equilibrium ratio of $N_{H^0}/N_{H^+}$ by considering the ionization balance from the known size, estimated $T_e$, and a Zanstra guess for the ionizing-photon supply rate [26]. The recent $HST$ spectra at H Ly $\alpha$ of I Zw 18 [38], provides an additional constraint on the Lyman continuum flux of about $2.9 \times 10^{51} \mathrm{s}^{-1}$. We use our illuminated slabs results together with $T_e$ and $N_e$ from Skillman & Kennicutt [25] to get: $N_{H^0}/N_{H^+} \approx 8 \times 10^{-4} f_2 f_1^{-1/2}$. Here $f_1$ and $f_2$ are volume filling factors (from 0.5 to 0.01 – [39]), accounting for clumping and other non-uniformities in the gas density and optical thickness. We find that even such small amounts of $H^0$ lead to a 1% to 2% effect on the derived helium abundance, via its effect on the emergent H$\alpha$ flux (see Table 3 in ref. [35]).

Though small, the error is systematic; always resulting in an increased Y. The case of I Zw 18 is given above for illustration, but in general additional knowledge of the geometry of the H II region are necessary in order to build a realistic model [40], and such is impossible to obtain for distant extragalactic H II regions.

A second process is incomplete absorption of hydrogen Lyman lines due to the finite optical thickness of the gas and dust, or photoionization of metals [36]. Balmer line emissivities are reduced relative to those for Case B. We calculate continuum and line opacities in order to use the results of Hummer & Storey [36] for $N_e = 100$ cm$^{-3}$. For I Zw 18 the effects due to dust are increasingly negligible due to its low dust-to-gas ratio [38]. Hence, we do not take them into account in our calculations of the systematics for this case.

Another effect, which we do not include here, is due to dust emission. We point out, however, that for all the $H$ lines of interest, it has a systematic deviation of the same sign as the effects already discussed. It is mostly significant for the radio recombination lines [36].

The overall systematic effect on the abundances based on H$\alpha$/(HeI $\lambda$ 6678) and $T_e$ in the 16,000–20,000 K are in the range of 1%–3%.

*[(2)]: The invalidity of Case B for He I:* In general, recombination theory is rarely simply valid for neutral helium, because of the distinct singlet and triplet species of He I, with the strongly metastable $2^3$S state of the triplets (Osterbrock [24], Ch.4.6). In the calculations of effective recombination coefficients the triplet series is often treated only as Case B, while the singlets are treated in both Case A and B (e.g. ref. [42] [43]). Therefore, whatever the detail and precision employed in such models, they do not account for important radiative transfer effects within He I, like the intersystem transitions coupling the singlet and triplet series (the latest models [43] include only those between n = 2 levels). Through this coupling the accurate calculation of the singlets of He I becomes dependent on the detailed treatment of the triplets, which are more sensitive to collisional and self-absorption effects in low-density plasmas. The use of simple recombination theory excludes also any consideration of inter-species radiative transfer effects. One example is the coupling between the hydrogen Lyman continuum ($\lambda$ 912) and the He I ($\lambda$ 504) continuum, which can become optically thick only in accord with the optical depth of hydrogen, $\tau_{912}$, because the dominant opacity at 504 Å is due to the Lyman continuum. Once the Lyman continuum becomes optically

thick, the ionization equilibrium of hydrogen shifts in favor of H I and affects $N_e$; hence − the total recombination rate of He I.

In order to illustrate the problem and derive some estimates on the systematics involved, we have calculated detailed models, extending to lower densities the models of Almog & Netzer [30]. In Figure 1 our results are compared to Case B calculations, and to two observed low-metallicity H II galaxies. This line-ratio diagram juxtaposes the singlets (He I $\lambda$ 6678) to the triplets (He I $\lambda$ 7065). Through the fluorescent enhancement of $\lambda$ 7065, this diagram can also indicate large optical depth, $\tau_{3889}$, in the He I $\lambda$ 3889 line. In the lowest density regime (1–100 cm$^{-3}$), both the singlet $\lambda$ 6678 line and the He I $\lambda$ 5876 line are not sensitive to $T_e$ or $N_e$, and their ratio is practically constant. This is not true for the triplet $\lambda$ 7065 transition which is fluorescent (coupled to $\lambda$ 3889), its emissivity increases with $\tau_{3889}$, as collisions and optical depth become more important.

Figure 1 illustrates the degree to which existing models account for the observed fluxes of He I lines used in the abundance determinations. No model fits the observations well but the Case B models fail most conspicuously to come close to the generous observational error boxes. The singlet line He I $\lambda$6678 is more reliable but still suffers from systematic effects. From Figure 1 it is also obvious that Case B models underestimate the lines sensitivity to $N_e$ (for both He I systems), as is inherent to Case B. Notice how in the detailed models the density sensitivity increases with $T_e$, as expected, while no such trend is discernible in the Case B models.

Despite the availability of detailed H II region photoionization models [40] the current quoted values of $Y_P$ are based on assumption of Case B and the emissivities of Brocklehurst [44]. The latter neglect the metastability of the $2^3$S state of He I, and are known to have other problems with the triplets fluxes [42], thus affecting the results of analyses which used mostly triplet He I lines to determine Y. Unfortunately, although small, the discrepancy is systematic (underestimating Y).

In summary, the systematic effect due to problem (2) – the invalidity of Case B for He I [44], would increase Y by 4%–6% (for He I $\lambda$ 6678) and 5%–8% (for He I $\lambda$ 4471) for $T_e$ between 10,000 and 20,000 K.

*[(3)] Neutral helium:* This problem arises from the lack of observational indication of the amounts of neutral helium (similarly to hydrogen) in an H II region. Yet any $N_{\text{He}^0}$ not accounted for underestimates the final abundance Y. Very detailed model calculations constrained by the best for any H II region observations, estimate $N_{\text{He}^0}$ at 3.2% in the Orion Nebula [45]. This nearby H II region is rich in metals and dust; dust extinction tends to keep ionization high and $N_{\text{He}^0}$ low. As we go to the extragalactic H II regions, the problem becomes extremely difficult to solve correctly and consistently, because of its strong dependence on geometry and composite H II regions [40]. The systematic error due to the presence of neutral helium can vary from zero to 3%. This follows from the range of values for the amount of neutral hydrogen (Problem 1) we obtain from the available parameters of extragalactic H II regions.

We sum up all the systematic effects, for each of the observations, taking into account the collisional corrections that have been made already while deriving each of the abundances, i.e. we do not double count systematics. The helium abundance derivations in the last two studies [22] [25] [41] in Table 1 are based on (the more reliable) He I $\lambda$ 6678 line and H$\alpha$, hence we consider their systematics. For the study by Pagel et al., we have considered also the systematic effects on He I $\lambda$ 4471, $\lambda$ 5876, and H$\beta$. The corrections for each $Y$ value due to each radiative transfer problem are listed in Table 1. The final upper bounds result from the systematic effects added, observers' collisional corrections subtracted, and the 2$\sigma$ observational statistical bound included. For each upper bound on $Y_P$ we derive a corresponding bound on the baryon-to-photon ratio, $\eta_{median}$ in terms of $\eta_{10} = \eta \times 10^{10}$, from the most recent theoretical relation of Kernan & Krauss [1] and their 95% confidence limits, $\eta_{95\%c.l.}$.

Table 1: The Upper Bounds on $Y_P$, $\eta$ in terms of $\eta_{10} = \eta \times 10^{10}$, and $\Omega_{BBN}$.

|  | Pagel et al. He vs. O | Pagel et al. He vs. N | Olive & Steigman | I Zw 18 NW |
|---|---|---|---|---|
| $Y_P \pm 1\sigma$ | 0.228 ±0.005 | 0.233 ±0.004 | 0.232 ±0.003 | 0.231 ±0.006 |
| systematics: Problem 1 Problem 2 Problem 3 | +0.005 +0.013 +0.002 | +0.005 +0.013 +0.002 | +0.006 +0.010 +0.002 | +0.005 +0.008 +0.002 |
| U.B. on: $Y_P$ | 0.257 | 0.260 | 0.255 | 0.258 |
| U.B. on: $\eta_{10,median}$ $\eta_{10,95\%c.l.}$ | 17.72 19.43 | 25.06 28.22 | 14.00 15.52 | 19.85 22.06 |
| U.B. on: $\Omega_{BBN}h^2$ $\Omega_{BBN}(h=0.5)$ $\Omega_{BBN}(h=0.7)$ | 0.071 0.284 0.145 | 0.103 0.412 0.21 | 0.0566 0.226 0.116 | 0.0805 0.322 0.164 |

### III. THE IMPLICATIONS FOR $\Omega$

From the above analysis we put most weight on two of the derived upper bounds: the one by Olive & Steigman, and the one of the NW H II region of I Zw 18 [25]. Their essence is different. While Olive & Steigman give a value of $Y_P$ based on careful extrapolation of the relation through the set of most recent H II region abundances, there is only a $Y$ value for I Zw 18 which is by no means truly primordial $Y_P$. However, $Y$ in I Zw 18 is very carefully derived from modern high-quality data by Skillman & Kennicutt [25], hence we are more confident about the size of the systematic errors (0.015) applied to it. The gas in I Zw 18 is not primordial, but is the closest to primordial we can get in H II regions [38], and it is good to keep its $Y$ value in mind. As a result of all this, we suggest that current $^4$He abundance derivation methods do not exclude a $Y_P \leq 0.255$, and perhaps even a value as high as $Y_P \leq 0.258$, see Figure 2. We would like to point out that $Y_P \leq 0.255$ is in agreement with the values and bounds for $^7Li$ and the current low limit on D and $D +^3 H_e$ (see section I).

We calculate an upper bound on $\Omega_{BBN}$ using the connection between $\eta_{10}$ and $\Omega_{baryon}$: $\Omega_{baryon} h^2 = 0.00365 \; \eta_{10} \; T_{2.726}^{-3}$ (here h is the Hubble constant in units of 100 and $T_{2.726}$ the temperature of the CMBR in units of 2.726 K). From the above limits on $Y_p$ we get that $\Omega_{BBN}$ can be as high as $0.22 \; h_{50}^{-2}$, and it might be even as large as $0.32 \; h_{50}^{-2}$. Our results using $\eta_{10, 95\% c.l.}$ are summarized in Table 1.

The new upper limits on $\Omega_{BBN}$ enable us to reconsider many cosmological models which were in conflict with the old limits. A universe with $\Omega_{total} = \Omega_{BBN} = \Omega_{dyn}$ is in agreement with the new bound and there is no need to invoke non-baryonic dark matter. One such a model is the PBI [4] model with $\Omega \approx 0.1$. Moreover the model that was suggested by Gnedin and Ostriker in 1992 [8] in which $\Omega_{BBN} = \Omega_{dyn}$ needed high values of $Y_P$. Since this model was inconsistent with observations at that time, Gnedin and Ostriker had to turn to unconventional scenarios in order to justify the high $Y_P$ values. These $Y_P$ values are well inside our new limit and therefore this model can rely on conventional cosmology. And finally a value of $Y_P$ near 0.251 is all that is needed in order that the fraction of baryons in clusters [3] will agree with the value of $\Omega_{BBN}$ in a flat universe dominated by cold dark matter.

## IV. CONCLUSIONS

In this paper we do not determine a new value for the primordial helium abundance, $Y_P$; we determine only the size of some important systematic errors. When accounted for, they provide an upper bound on $Y_P$ which is less confining than currently assumed. The new bound on $Y_P$ provides an upper limit to $\Omega_{BBN}$ which is much less restrictive for cosmological models.

The new relatively high upper bound on $\Omega_{BBN}$ removes the need to turn to exotic models in order to reconcile between various observations of the density parameter. Moreover, the simplest and most esthetic model in which $\Omega_{total} = \Omega_{BBN} = \Omega_{dyn} \approx 0.1$, a model that does not rely on unobserved entities (dark matter) or unconventional physics and is consistent with many observations ( [46] and references there), is also consistent with BBN limits.

Because of the logarithmic dependence of $Y_P$ on $\eta$, a very accurate determination of $Y_P$ is needed in order to determine $\Omega_{BBN}$ precisely. Unfortunately, because of the observational limitations of the H II galaxy method, all we were able to calculate in this paper are upper bounds on $Y_P$ and, therefore, we are unable to constrain further the cosmological models. Two of the important sources of systematic errors in the H II galaxy method discussed here are related to the geometry and stellar content of the H II regions. Such constraints cannot be achieved with the resolving power of any current telescope. Even with the best-observed nearest H II region (the Orion Nebula) the uncertainties in the helium abundance are still difficult to reduce below 7% [45]. Therefore, modern sophisticated radiative transfer techniques cannot be applied with equal success to the extragalactic H II regions. As an alternative, extreme halo stars in our Galaxy may provide a way to derive $Y_P$ with precision interesting for cosmology. Much of the improvement will come from the fact that halo dwarfs are much closer to primordial abundances in their unprocessed atmospheres. This is illustrated in Figure 2, where some of the program halo stars are marked with respect to the H II regions of the lowest known metalicity. A project to derive helium abundances in extreme halo dwarfs and obtain an independent and more accurate value of $Y_P$ is underway [33] [47]. Once we can determine accurately the value of $Y_P$ we will be able to reject/verify the validity of the various cosmological models.

*Acknowledgments.* We are thankful to D. Proga and P. Kernan for providing us with results prior to publication.

## V. FIGURE CAPTIONS

**Figure 1**: Comparison between observed and calculated He I intensity ratios, on a logarithmic scale, for the H II region in UGC 4483 ($T_e$=16,600 K, $N_e \approx 100$ cm$^{-3}$) and the H II galaxy II Zw 40 (12,700 K, 225 cm$^{-3}$). The tick marks along the curves of the detailed models correspond to densities (from left to right) of 1, 10, 100, $10^3$, $10^4$, and $10^5$ cm$^{-3}$. The solid line is for the model with $T_e$=20,000 K. The plotted model of Almog & Netzer [30] was specially designed for comparison to Brocklehurst [44]. Note that dust extinction, $A_V$, would shift the observed line ratios towards the upper right corner and further away from the models.

**Figure 2**: $Y$ versus oxygen abundance (logarithmic scale relative to solar abundances) for the extragalactic metal-poor H II regions of Pagel et al., and the recently analyzed I Zw 18 and UGC 4483. Such diagrams are used for extrapolation to zero metallicity (in this case, oxygen) to derive $Y_P$. The recent value of $Y_P = 0.232$ by Olive & Steigman is shown together with our upper bound estimated here; the same for the recent $Y$ abundance of I Zw 18. The marks above the abscissa are some of the halo dwarfs from the project to obtain an accurate value of $Y_P$.

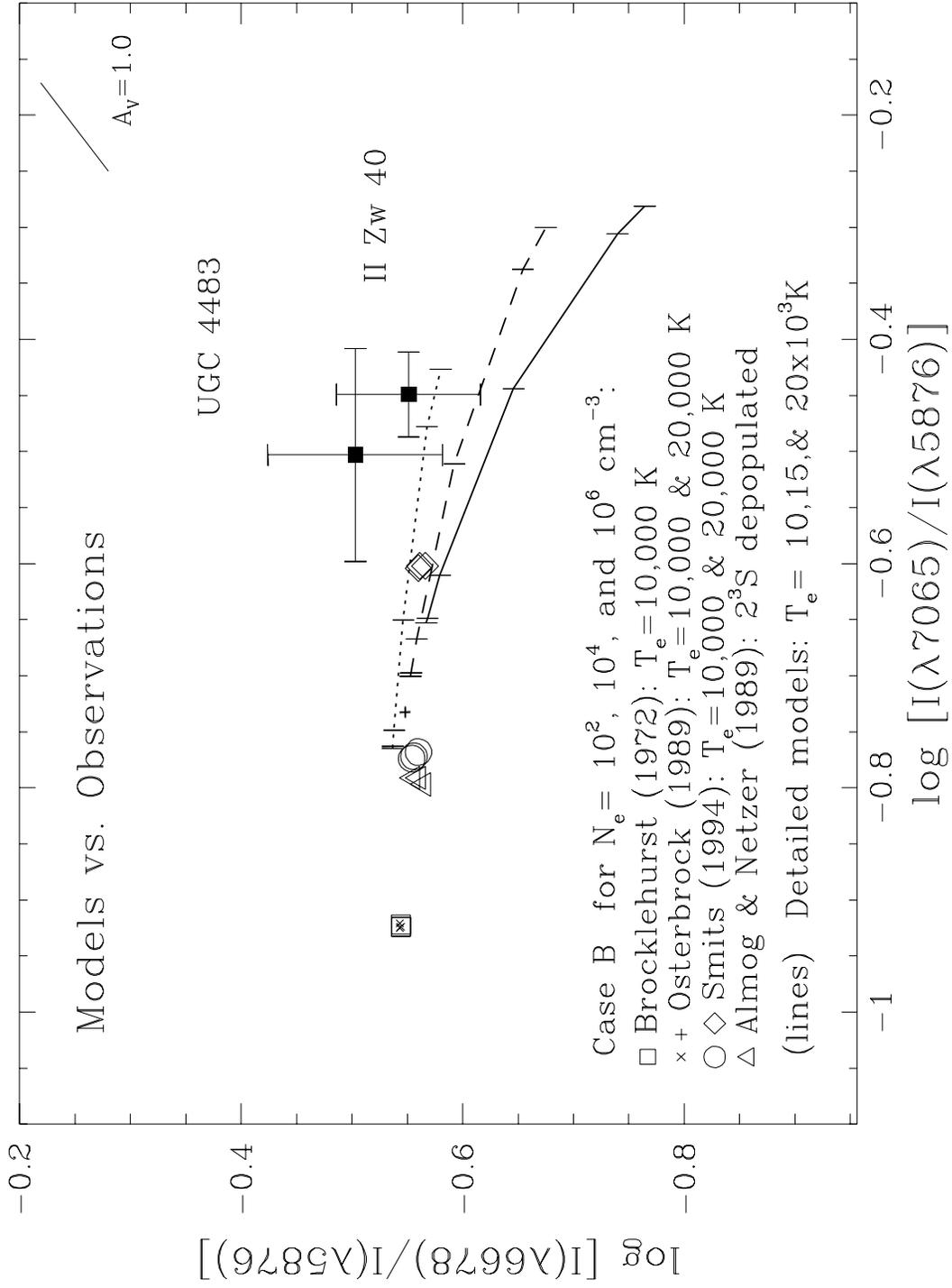

Extragalactic H II regions
Pagel et al.(1992):  x & fit
S&K (1993):  o
.... Our upper bounds

Helium Abundance, Y

Oxygen Abundance, [O/H]